\documentclass[preprint2]{aastex}

\usepackage{graphicx} 
\usepackage{figcaps}
\usepackage{balance}
\usepackage{aastex_plus}

\shorttitle{Combining Density Diagnostics and DEM to Constrain Steady Heating}
\shortauthors{Winebarger et al.}

\figcapsoff

\begin{document}


\title{Using a Differential Emission Measure and Density Measurements in an 
Active Region Core to Test a Steady Heating Model}

\author{Amy R.\ Winebarger}
\affil{NASA Marshall Space Flight Center, VP 62, Huntsville, AL 35812}
\email{amy.r.winebarger@nasa.gov}
\and
\author{Joan T.\ Schmelz}
\affil{Physics Department, University of Memphis, Memphis, TN 38152}
\and
\author{Harry P.\ Warren}
\affil{Space Science Division, Naval
       Research Laboratory, Washington, DC 20375}
\and
\author{Steve H.\ Saar \and Vinay L.\ Kashyap}
\affil{Harvard-Smithsonian Center for Astrophysics, 60 Garden St., Cambridge, MA 
02138}


\begin{abstract}
The frequency of heating events in the corona is an important constraint on the 
coronal heating mechanisms. Observations indicate that the intensities and 
velocities measured in active region cores are effectively steady, suggesting 
that heating events occur rapidly enough to keep high temperature active region 
loops close to equilibrium. In this paper, we couple observations of Active 
Region 10955 made with XRT and EIS on \textit{Hinode} to test a simple steady 
heating model. First we calculate the differential emission measure of the apex 
region of the loops in the active region core. We find the DEM to be broad and 
peaked around 3\,MK. We then determine the densities in the corresponding footpoint 
regions. Using potential field extrapolations to approximate the loop lengths and 
the density-sensitive line ratios to infer the magnitude of the heating, we build 
a steady heating model for the active region core and find that we can match the 
general properties of the observed DEM for the temperature range of 6.3 $<$ Log T 
$<$ 6.7. This model, for the first time, accounts for the base pressure, loop 
length, and distribution of apex temperatures of the core loops. We find that the 
density-sensitive spectral line intensities and the bulk of the hot emission in 
the active region core are consistent with steady heating.  We also find, however, 
that the steady heating model cannot address the emission observed at lower 
temperatures. This emission may be due to foreground or background structures, 
or may indicate that the heating in the core is more complicated. Different 
heating scenarios must be tested to determine if they have the same level of agreement.
\end{abstract}


\section{Introduction}

A major problem in coronal physics is to determine the mechanisms that 
transfer and dissipate energy that heat the million-degree corona.  Regardless 
of the transfer or dissipation mechanism, the predicted timescale for energy 
release for all mechanisms is finite, i.e., a single heating event is relatively 
short-lived \citep{klimchuk2006}.  The {\em frequency} of heating events 
occurring on a single strand in the corona, however, could distinguish between 
the different theoretical mechanisms.  (Here, we use the term ``strand'' to 
refer to the fundamental flux tube in the corona, and the term ``loop'' to refer 
to a discernable structure in an observation.  A loop could be formed of a 
single strand, which would imply we are currently resolving the corona, or it 
could be formed of many, sub-resolution strands.)

There has been significant debate as to whether observations support high- or 
low-frequency heating.  If heating events that occur on a single strand are 
frequent (i.e., the time between individual heating events is much shorter than 
the cooling time of the plasma) and approximately uniform, then the temperature 
and density of the plasma along the strand are effectively steady as the plasma 
does not have time to cool or drain between heating events.  Such a heating 
scenario predicts a loop with steady intensities and velocities.  If the heating 
events are infrequent (i.e., the time between heating events is much longer than 
the cooling time of the plasma), then the resulting temperature and density of 
the plasma along the strand will evolve; such a heating scenario is often termed 
``nanoflare heating'' \citep[e.g.,][]{cargill1997}, though it could be 
representative of different types of mechanisms, not simply Parker's canonical 
nanoflare theory \citep{parker1972}.  The evolution of the plasma may or may not 
be apparent in loop observations depending on the number of strands the make up 
the loop and the relative timing of their heating events.  In the case of a loop 
formed of only a few strands that undergo a burst of heating events close in 
time, the observed loop's properties, such as the intensity of the loop in a 
spectral line or filter, would appear to evolve as the strands evolve 
\citep{warren2002b}; such a heating scenario is sometimes called a ``short 
nanoflare storm'' \citep{klimchuk2009}.  If the loop is formed of many 
sub-resolution strands, each strand being heated and evolving randomly, then the 
observed loop's intensity would appear steady regardless of the dynamic nature 
of the plasma along a single strand; this type of low-frequency heating is 
sometimes called a ``long nanoflare storm'' \citep{klimchuk2009}.

Observations in active regions show both evolving and steady structures 
\citep{reale2010}.  So-called ``warm,'' 1\,MK loops that form a bright arcade in 
EUV images are evolving and their properties are consistent with low-frequency 
heating in the form of short-nanoflare storms 
\citep{winebarger2003a,warren2003,winebarger2005,ugarte2006,ugarte2009,mulu2011}.  
The intensities of the hot ($>$ 2\,MK) loops that make up the active region 
cores, however, appear to be steady over many hours of observation  
\citep[e.g.,][]{warren2010a,warren2011}.  This steadiness is apparent in both 
the X-ray intensity over the neutral line and in the footpoints of the hot loops 
observed in the EUV and commonly called moss \citep[e.g.,][]{antiochos2003}.   
Additionally, the velocities and non-thermal velocities in the moss are also 
steady for hours of observations \citep{brooks2009}.  The steadiness of the 
intensities and velocities has prompted many to hypothesize the heating in the 
core regions is effectively steady.  However, different heating scenarios, such 
as the long nanoflare storm scenario described above, also predict steady 
intensities and velocities.  Additional comparisons between observations and 
simulations are necessary to discern whether the active region core is 
consistent with effectively steady heating.  \cite{warren2011} demonstrated that 
the properties of the plasma along strands suffering effectively steady 
(high-frequency) heating can be well approximated by the steady heating 
solutions to the hydrodynamic equations.  In this paper, we will determine 
whether observations of the loops that form the active region core are 
consistent with the predictions of steady solutions to the hydrodynamic 
equations.

If the density and temperature in a well-isolated and resolved loop was known, 
it would be possible to test whether the loop was heated steadily or not using 
the so-called Rosner-Tucker-Vaiana (RTV) scaling laws (\citealt{rosner1978}, also 
see \citealt{serio1981}).  These scaling laws define a relationship between the 
apex temperature, $T_{\rm apex}$, base pressure, $p_0$, and loop half-length, 
$L$, for steady uniform heating, i.e., $T_{\rm apex} = 1.4 \times 10^3 (p_0 
L)^{1/3}$. However, individual loops are very difficult to isolate in active 
region cores, and loops that appear to be resolved may themselves be formed of 
sub-resolution strands, hence such a direct comparison with model predictions is 
not possible. 

Another method to test whether a heating scenario is viable is to forward 
model the active region or full Sun as ensembles of strands with either a steady 
or infrequent heating model and then compare simulated images from the model to 
the observed images.  Several studies have been completed using this technique 
(\citealt{schrijver2004,warren2006,warren2007,lundquist2008a}).  
In these studies, the field lines from the potential magnetic field were used to 
approximate the global or active region structure.  The heating rate was assumed 
to be a function of the magnetic field strength and loop length.  
Several different parameters were varied, such as the functional form of the 
heating, the degree in which the loops expanded, and the non-uniformity of the 
heating along the loop.  These studies determined that to best match the X-ray 
observations of the active regions, the heating must be proportional to average 
magnetic field strength and inversely proportional to loop length, the loops 
must expand with height, and the heating along the loop must be near uniform.  
(Note that loop expansion contradicts the observation that loops have near 
constant cross sections along their lengths; \citealt{klimchuk2000}.)  
There are two significant shortcomings of these previous studies.  First, most 
of the studies relied on matching the intensities in a single X-ray filter.  
Second, the EUV footpoint emission from the moss was poorly matched by the 
models; in general, the moss emission was too bright.

In an attempt to simultaneously match both the EUV moss emission and the apex 
X-ray emission in two filters, \cite{winebarger2008} modeled an active region 
core as an ensemble of strands.  Instead of assuming a functional form of the 
heating, however, they used the \textit{TRACE} 171\,\AA\ intensity in the moss 
region to limit the steady heating rate.  The \textit{TRACE} 171\,\AA\ intensity 
is pressure sensitive given an unknown filling factor \citep{martens2000}.  In 
this study, a filling factor was assumed for the entire moss region, then the 
pressure was derived from the TRACE 171\,\AA\ image.  For each strand, a heating 
rate was determined that would match the pressure; the expected intensities in 
the X-ray filter data was then calculated.  This process was completed for 
several different assumed filling factors.  The best agreement for the X-ray 
filter intensities was found for loops that expand inversely with the magnetic 
field strength and for a filling factor of 8\%.  This study was the first 
successful steady heating model to account for both the footpoint EUV emission 
and apex X-ray emission.  The weakness of this study was the forced assumption 
of a single filling factor for the entire moss region.

In this paper, we will determine whether observations from \textit{Hinode} XRT 
and EIS agree with a steady heating model of the active region core.  We build 
upon the previous full models of active region cores with two key improvements.  
First, instead of comparing model intensities with observed intensities in one 
or two filter images, we compare a differential emission measure (DEM) from the 
model with a DEM calculated from a combination of XRT filter and EIS spectral 
line intensities.  Second, we use two density sensitive spectral lines 
(\ion{Fe}{12} 186.880/195.119\,\AA) in the moss region to simultaneously 
constrain both the pressure in the core loops and the filling factor.  As with 
the previous studies, we use potential field extrapolations to approximate the 
loop length and geometry, assume the loops expand with height and assume the 
energy is deposited uniformly along the loop.  We determine that the DEM from 
the steady heating model well matches the DEM from observations around the peak 
temperature in the active region core (6.3 $<$ Log T $<$ 6.7), as well as the 
spectral line intensities in both the \ion{Fe}{12} 186.880 and 195.119\,\AA\ 
lines in the footpoints of the active region.  However, the steady heating model 
does not reproduce the observed DEM in the ``warm'' temperature range (6.1 $<$ 
Log T $<$ 6.3).  This warm emission could originate from the overlying arcade 
and not from the core loops, or it could indicate that the heating in the active 
region core is more complicated.  These results provide, for the first time, 
evidence that the high temperature loops in the core of an active region and 
their footpoint pressure can be well represented by high-frequency heating.  Different 
heating scenarios must be tested to determine if they have the same level of 
agreement.  


\section{Data}

Active Region 10955 was observed on 2007 May 13 by \textit{Hinode} and 
\textit{Solar Heliospheric Observatory (SoHO)} instruments.  A full disk X-ray 
Telescope image is shown in Figure \ref{fig:fulldisk}. The observations 
considered in this analysis were taken between 14:25 and 16:30 UT.  At this 
time, the active region was approximately \{445'',-149''\} away from disk 
center.

The X-Ray Telescope (XRT; \citealt{golub2007}) on \textit{Hinode} 
\citep{kosugi2007} is a broadband instrument similar to the \textit{Yohkoh} Soft 
X-Ray Telescope \citep{tsuneta1991}, but with better spatial resolution, 
sensitivity, and temperature coverage. XRT has 1\arcsec pixels, 2\arcsec 
resolution, and 9 temperature sensitive filters that can be used alone or in 
combination. The XRT data used for this analysis were taken on 2007 May 13 from 
16:17 - 16:30 UT. The data set is comprised of full Sun, full resolution images 
in multiple filters and filter combinations. The data processing included 
dark-frame subtraction, vignetting correction, and high-frequency noise removal using 
the standard \verb+xrt_prep+ routine available from SolarSoft.  Spacecraft jitter 
was removed using \verb+xrt_jitter+, and long and short exposures (see Table 
\ref{tab:xrtdata}) were co-aligned and combined to increase the image dynamic 
range. Additional Fourier filtering was done to the data from the thickest 
channel, Be\_thick, to remove low-level, residual, longer-wavelength noise 
patterns. We have used updated filter calibrations (\citealt{narukage2011}) and 
accounted for a 1640-\AA\ thickness of the time-dependent contamination layer on 
the CCD.  (We used the ``best-model'' contaminant composition of diethylhexyl 
phthalate, as provided by the XRT calibration software, but we note that the 
true composition is as of yet undetermined.)  

The EIS instrument on \textit{Hinode} is a high spatial and spectral resolution 
imaging spectrograph. EIS observes two wavelength ranges, 171--212\,\AA\ and 
245--291\,\AA, with a spectral resolution of about 22\,m\AA\ and a spatial 
resolution of about 1\arcsec\ per pixel. There are 1\arcsec\ and 2\arcsec\ slits 
as well as 40\arcsec\ and 266\arcsec\ slots available.  Solar images can be made 
using one of the slots or by stepping one of the slits over a region of the Sun. 
Telemetry constraints generally limit the spatial and spectral coverage of an 
observation. See \cite{culhane2007} and \cite{korendyke2006} for more details on 
the EIS instrument.

In this analysis we consider EIS observations taken from 14:25:57 to 15:15:10 UT 
on 2007 May 13. For these observations the 1\arcsec\ slit was stepped across the 
active region and a 10\,s exposure was taken at each position. The area observed 
was $256\arcsec\times256\arcsec$. A total of 16 data windows, some containing 
multiple emission lines, were downloaded from the spacecraft. The raw data was 
processed to remove the CCD pedestal, dark current, and spurious intensities 
from warm pixels. The pre-flight calibration was also applied to the data. 
Before the line profiles are fit it is necessary to correct for an oscillation 
in the EIS wavelength dispersion. We do this by assuming that there are no net 
velocity shifts along the bottom 30 pixels of the slit (see \citealt{brown2007} 
for details). For each emission line of interest the best-fit parameters for a 
single Gaussian were calculated.  
 
The XRT and EIS data were aligned by cross correlating the Be\_thin filter image 
from XRT with the \ion{Fe}{16} 262.984\,\AA\ raster from EIS.  The EIS field of 
view is shown in the XRT full disk image in Figure \ref{fig:fulldisk} as solid 
white lines.  The entire EIS field of view for the C\_poly filter and 
\ion{Fe}{12} 195.119\,\AA\ raster are also shown in Figure \ref{fig:fulldisk}.  

We use magnetic field measurements from the Michelson Doppler Imager (MDI) on 
\textit{SoHO}.   All of the 96-minute full disk magnetograms taken within 5 
hours of the EIS data were aligned and averaged to reduce noise.  The MDI 
magnetograms were aligned to the \ion{Fe}{12} 186.880\,\AA\ data.  The averaged 
MDI data in the EIS field of view with the \ion{Fe}{12} 186.880\,\AA\ contours 
is shown in Figure~\ref{fig:fulldisk}.

\section{Analysis}

\subsection{Differential Emission Measure}

The first goal of this research is to determine the distribution of emission 
from the apex of the core loops.  To complete this goal, a region between the 
moss in the core of the active region was selected.  This region is shown in 
Figure \ref{fig:multipanel} with solid lines in several example rasters or 
filter images.  The emission in this region represents the intensity from the 
core of the active region plus ambient corona between the Sun and the telescope.  
To remove the ambient corona, we selected four background regions at the edge of 
the field of view, shown in Figure~\ref{fig:fulldisk} as dashed lines.  We 
average the intensity in the region of interest and subtract the average 
background intensity.  The background-subtracted intensities are given in 
Table~\ref{tab:intensity}.  EIS intensities are given in ergs cm$^{-2}$ s$^{-1}$ 
sr$^{-1}$ and XRT intensities are given in DN s$^{-1}$ pix$^{-1}$.  

For this work we have included EIS observations of the \ion{Fe}{17}
line at 254.87\,\AA. This line is problematic for several
reasons. Both the emissivity of this line and the effective area of
EIS at this wavelength are relatively small, which makes the line
difficult to observe. Furthermore, the atomic data for this transition
appears to be inconsistent with the other \ion{Fe}{17} and high
temperature Ca lines observed with EIS \cite{warren2008}. These
differences appear to be about a factor of 2. Since this is the
highest temperature line observed with EIS outside of flares it is
useful to accept these limitations so that we can have better overlap
between the EIS and XRT observations.

The uncertainties in the EIS intensities given in Table~\ref{tab:intensity} are the calculated 
statistical errors due to photon noise and the error in the Gaussian fits and
an assumed 22\% systematic error associated with the absolute calibration of 
the EIS data \citep{lang2006}.  The uncertainties in the XRT intensities due to photon 
noise are more difficult to assess.  The method used to calculate the XRT errors 
in Table~\ref{tab:intensity} is described below.

To generate possible DEM curves that can reproduce the observed fluxes given in 
Table \ref{tab:intensity}, we have used \verb+xrt_dem_iterative2+, which was 
designed originally for use with XRT data only but has been modified slightly to 
allow for inclusion of EIS data as well. Leading up to the launch of 
\textit{Hinode}, \cite{golub2004} and \cite{weber2004} tested and validated the 
method with synthetic data. The routine employs a forward-fitting approach where 
a DEM is guessed and folded through each response to generate predicted fluxes. 
This process is iterated to reduce the $\chi^2$ between the predicted and 
observed fluxes. The DEM function is interpolated using several spline points, 
which are directly manipulated by \verb+mpfit+, a well-known and much-tested IDL 
routine that performs a Levenberg-Marquardt least-squares minimization.  This 
routine uses Monte-Carlo iterations to estimate errors on the DEM solution. For 
each iteration, the observed fluxes in each filter were varied randomly within the 
uncertainties and the program was run again with the new values.

The \verb+xrt_dem_interative2+ program requires user input for the XRT response 
functions and EIS emissivity functions.   The XRT filter responses were 
calculated using the XRT standard software (\verb+make_xrt_wave_resp+ and 
\verb+make_xrt_temp_resp+).  These programs account for the time dependent 
contamination of the XRT CCD.  The EIS line emissivity functions were calculated 
using CHIANTI 6.0.1. The default abundances (\citealt{feldman1992}) and 
ionization equilibrium (\citealt{mazzotta1998}) were used.  The emissivity 
functions for some ions are density sensitive.  As a default, we find a density 
from the \ion{Fe}{13} 202.044/203.826\,\AA\ intensity ratio.  The ratio of the 
two background subtracted intensities given in Table~\ref{tab:intensity} return 
a density of Log $n_e$ = 9.7 cm$^{-3}$.  All emissivity functions were calculated 
using this density.

The data used in this analysis were taken on 2007 May 13, just before the first 
XRT CCD bake-out (2007 July 23). This bake-out was designed to remove the 
accumulating wavelength-dependent contamination layer that was affecting the 
instrument sensitivity. As a result, the contamination layer was close to 
maximum thickness. In addition, the XRT team was not yet taking regular G-band 
images, which is one of the tools used to estimate the thickness of the 
contamination layer. As a result, the thickness is not as well known.  Al\_mesh 
observations are generally the most affected by contamination, especially at low 
temperatures where the sensitivity is dependent mainly on longer wavelength 
spectral lines. In light of these uncertainties, and combined with the fact that 
we have EIS spectral lines that effectively cover the lower coronal 
temperatures, we have elected to eliminate Al\_mesh from the DEM calculation. 

XRT is a broadband instrument allowing photons of many different energies to 
generate electrons on the detector.  The number of electrons that are generated 
are proportional to the photon energy, so there is no {\it a priori} way to 
deduce the number of photons that produced a signal from the number of electrons 
deposited onto the detector.  The measured count rate in a pixel could be from a 
few high-energy photons or from many low-energy photons, and the uncertainty 
would vary accordingly.   Below we describe how we use a ``bootstrap'' method to 
calculate the errors associated with photon uncertainties.  

To determine the XRT filter errors due to photon noise, we first calculate a DEM 
using the method described above assuming a 20\% error in the XRT filter 
intensities.  We then use that DEM to calculate the emerging spectrum from the 
solar plasma, $I_{solar}(\lambda)$, in photons s$^{-1}$ cm$^{-2}$ sr$^{-1}$ 
\AA$^{-1}$.  We convolve the spectrum with the effective area for each filter 
and multiply by the resolution, wavelength bin size, and exposure time for each 
filter to find the spectrum in photons at the detector in a given filter, i.e., 
\begin{equation}
I_{det}(\lambda) = 
I_{solar}(\lambda)*EA_{filter}(\lambda)*\Delta\lambda*t_{exp}*C.
\label{eqn:inten1}
\end{equation}
In the above equation, the effective area, $EA_{filter}$, is in cm$^{2}$ and is 
taken from the XRT program \verb+make_xrt_wave_resp+, the $\Delta\lambda$ is the 
size of the wavelength bin in \AA, $t_{exp}$ is the exposure time, and the 
factor, $C$, is the sterradians subtended by one XRT pixel.  After calculating 
the photons that interact with the detector, we determine the photon noise by 
taking the square root of this intensity, $\sigma_{det} = \sqrt{I_{det}}$.  The 
intensity in a given pixel in data number (DN) can then be found by  
\begin{equation}
I_{DN} = \sum I_{det}(\lambda)*\frac{E(\lambda)}{\epsilon_0} \frac{1}{G}
\label{eqn:inten2}
\end{equation}
where $E$ is the energy associated with an incoming photon, $E(\lambda) = 
hc/\lambda$, $\epsilon_0$ is the energy per electron on the detector (3.65 eV), 
and $G$ is the gain of the detector (59 electron DN$^{-1}$).  We then propagate 
the photon noise through this step to determine the error in the intensity in 
DN.  Note there are additional sources of statistical uncertainty in this calculation, but 
they are small compared to the photon noise.
We determine the relative uncertainty due to photon noise
in the XRT filters are $<$5\% in the thinnest filters and $>100$\% 
in the thickest filter.  We combine this statistical uncertainty with an assumed
20\% systematic uncertainty in the absolute calibration of the XRT intensities.
The uncertainties given in Table~\ref{tab:intensity} reflect 
this analysis. 
 
After completing the XRT error analysis, a new DEM was calculated with the 
updated errors.  The results from the DEM calculations are shown in the left 
panel of Figure \ref{fig:dem}.  The solid thick line shows the best DEM 
calculated for input intensities.  The DEM is broad with a peak at Log T = 6.5.  
The intensities calculated in each spectral line or filter for this DEM are 
given in Table~\ref{tab:intensity}.  In general, the ratio of the observed to 
modeled intensity is close to 1 with two exceptions.  The \ion{Fe}{17} spectral line
intensity is off by a factor of 2; this is likely due to poor atomic data and
is consistent with the results in \cite{warren2008}.
Additionally, the Be\_thick intensity calculated 
from the DEM is roughly a factor of 3 higher than observed.  

The dotted lines clustered around the solid thick line are DEMs calculated by 
varying the input intensities within the uncertainties and hence provide an 
estimate of the uncertainty of the DEM. Another way to assess the uncertainty in 
the DEM and determine the temperature regime where the DEM is well constrained 
is to calculate how much additional emission could be added to a single 
temperature bin without changing the modeled intensities in any spectral line or 
filter by more than the expected errors.  This value is shown as a blue line in 
Figure~\ref{fig:dem}.  For instance, the DEM calculated for Log T = 7.0 MK is 
approximately $6 \times 10^{14}$ cm$^{-5}$ K$^{-1}$.  We determine we could increase the 
DEM to $2 \times 10^{20}$ cm$^{-5}$ K$^{-1}$ (a factor of $3 \times 10^5$) without changing 
the modeled intensities by more than the errors. This implies this temperature 
bin is not well constrained and emission could be in this temperature bin, but 
the spectral lines and filters we are using in this analysis are not sensitive 
to it.  The ratio of this ``maximum DEM'' to the calculated DEM is less than 3.0 
in the temperature range of 6.1 $\leq$ Log T $\leq$ 6.7; we consider this 
temperature range well constrained by the observations. 

If we define the differential emission measure to be $\xi = n^2 ds/dT$, we can 
write the intensity in a given spectral line or filter, $I_\lambda$, in terms of 
the emissivity function or filter response function, $\epsilon_\lambda(T,n)$, 
i.e., 
\begin{eqnarray}
I_\lambda & = &  \frac{1}{4\pi} \int \epsilon_\lambda(T,n) \xi dT \\
1 & = & \int \frac{\epsilon_\lambda(T,n)}{4\pi I_\lambda} \xi dT\\
1 & = & \int \frac{1}{\xi_{\rm loci}(T)} \xi dT
\label{eq:emloci}
\end{eqnarray}
where we have introduced the emission measure loci function,
\begin{equation}
\xi_{\rm loci}(T) = \frac{4 \pi I_\lambda}{\epsilon_\lambda(T,n)}
\end{equation}
\citep{jordan1987}.  In the right panel of Figure \ref{fig:dem}, the  
emission measure distribution, i.e., $\xi(T)dT$, is shown in black.  The 
EM loci curves for the EIS lines considered in this analysis are shown in red 
and the EM loci for the XRT filter intensities are shown in green.  The blue 
line shows the maximum emission possible in a single bin without changing the 
modeled intensities by more than the errors.  Note that the thickest XRT filter 
(Be\_thick) does not well constrain the emission measure at high temperatures due to the 
large uncertainties in the intensities. 

\subsection{Moss Density}

In this research we use the density sensitive line intensities in the moss 
to constrain the steady heating model.  Though we use the actual intensities in 
the model, we present here a calculation of the density from the line ratio.  
This data set includes two density sensitive line pairs, \ion{Fe}{12} 
186.880/195.119\,\AA\ and \ion{Fe}{13} 203/202\,\AA.  We choose to use the 
cooler of these (\ion{Fe}{12}) to determine the density in the moss.   

The top two panels of Figure \ref{fig:density} show the two \ion{Fe}{12} 
spectral lines.  First we use the intensity in the density sensitive line, in 
this case \ion{Fe}{12} 186.880\,\AA, to define the moss.  We choose a threshold 
of 1200~ergs cm$^{-2}$ s$^{-1}$ sr$^{-1}$.  The moss regions are shown with 
contours in Figure~\ref{fig:density}.   

Before we calculate the density, we first need to subtract the background 
emission from the moss regions.  Because moss forms the footpoints of the high 
temperature loops, we are not only looking through the ambient corona, but also 
through the hot loops above the moss.  To account for this, we choose to use the 
central region of the core (shown as a rectangle in Figure~\ref{fig:multipanel}) 
as the background.  These background intensities are the same as the apex 
intensities and given in Table~\ref{tab:intensity}.

Using the ratio of the background subtracted intensity and the density sensitive 
emission measure ratio calculated from CHIANTI 6.0.1, we calculate a density for 
each moss pixel.  The lower left panel shows a density map of the moss regions.  
The density does not appear to depend on spatial location in the primary moss 
regions, though the satellite regions to the right typically have lower 
densities. The lower right panel is a histogram of the number of pixels in the 
moss region with a given density.   The average density is 
Log n = 10.15 cm$^{-3}$ and the standard deviation is 0.49.  The largest 
densities measured in this region are $\sim 5 \times 10^{10}$\,cm$^{-3}$.

\subsection{Loop Length}

We approximate the loop lengths and geometries in the active region core using a 
potential field extrapolation of the photospheric field.  First, we start with 
the full Sun, time-averaged photospheric magnetic field measurements, shown in 
the upper left panel of Figure~\ref{fig:magfield}.  We extract a region of full 
disk magnetic field around the active region, shown with an ``X'' on the figure.  
The extracted region, shown in the upper right panel of 
Figure~\ref{fig:magfield}, has been generated so that the pixels are 
approximately square with the pixel sizes measured in Mm from the center of the 
active region.  Hence, we create a magnetic field coordinate system that is 
Cartesian around the middle of the active region.  We transform the coordinates 
from the magnetic field coordinate system to the image coordinate system using 
the transformation matrixes in \cite{aschwanden1999}.  

After correcting the magnetic field, we numerically solve the equations 
$\nabla \times B = 0$ and $\nabla \cdot B = 0$  to determine the magnetic field 
vectors in the volume above the photospheric field (e.g., \citealt{gary1989}). 
To define the footpoints of the core loops, we transform the moss footpoints, 
originally in the image coordinate system, to the magnetic field coordinate 
system.  Recall that the original moss region was co-aligned with the original 
magnetic field before the region of interest was extracted.  Contours of the 
moss region in the magnetic field coordinate system are shown on the magnetic 
field image in the top right of Figure~\ref{fig:magfield}.  We then trace field 
lines from the moss regions.  The x,y coordinates of the moss are defined by the 
\ion{Fe}{12} 186.880\,\AA\ intensity.  We assume the height of the moss, and the 
originating z position of the field line, is 2.5 Mm above the photospheric 
field.  Figure~\ref{fig:magfield} shows a subset of the field lines projected 
onto the \ion{Fe}{12} 186.880\,\AA\ image (lower left panel), as well as a 
histogram of the full loop lengths of all the field lines in Mm (lower right 
panel).  We find the longest loops associated with this moss region are $<
100$\,Mm.

\section{The Steady Heating Model}

In this section, we test whether a simple, steady heating model of this active 
region core can reproduce both the density-sensitive spectral line intensities 
in the moss regions and the DEM of the apex emission from the core loops.  
We predicate this model on four key assumptions derived from the results of 
previous successful active region models. 1) We assume the geometry of the 
active region core is well represented by field lines from the potential field 
extrapolation.  2) We assume the atomic physics of the plasma is well 
represented by the CHIANTI 6.0.1 package with the ionization balance described 
by \cite{mazzotta1998} and abundances described by \cite{feldman1992}. 3) We 
assume the heating along the loop is well approximated by uniform, steady 
heating. 4) We assume the cross-sectional area of the loop increases as the 
magnetic field along the loop decrease, i.e., $A(s) \sim 1/B(s).$ 

Below we describe the process of finding the best solution for a single field 
line.  We complete this process for each field line, then derive a DEM from the 
portion of all the field lines that projects into the region of interest shown 
in Figure~\ref{fig:multipanel}.  To avoid duplicate field lines, we consider 
field lines traced from only the positive polarity magnetic field.  

The example field line is shown as a thick black line in the lower left panel of 
Figure~\ref{fig:magfield}.  The field line is 36.7 Mm long.  The (x,z) and (y,z) 
projections of the field line are shown in Figure~\ref{fig:example}.  These 
figures demonstrate that the field line starts and terminates 2.5 Mm above the 
solar surface and is roughly semi-circular in the x-projection.  The 
y-projection shows that the field line is slightly inclined.  We associate the 
field line with the background subtracted \ion{Fe}{12} 186.880 and 195.119\,\AA\ 
intensities at its originating footpoint, for this field line, 819.5 and 1213.2 
ergs cm$^{-2}$ s$^{-1}$ sr$^{-1}$, respectively. 

The first step in formulating the model is guessing the uniform heating rate 
that will match the \ion{Fe}{12} intensities at the base.  Recall that 
\cite{rosner1978} determined scaling laws that related the volumetric heating 
rate to the loop half length in cm and base pressure in dyn cm$^{-2}$, i.e.,
\begin{equation}
E = 1 \times 10^5 p_0^{7/6}/L^{5/6}.
\end{equation}   
Using the \ion{Fe}{12} line ratio, we calculate a density of $1.0 \times 
10^{10}$ 
cm$^{-3}$.  This density is determined assuming the line intensity is generated 
by plasma at the temperature of peak formation for \ion{Fe}{12}, Log T = 6.1.  
We can estimate the base pressure, then, by $p_0 = 2n_0kT$ where $k$ is 
Boltzmann's constant.  For this density, we estimate a base pressure of 3.6 dyn 
cm$^{-2}$.  Using this base pressure and the loop half-length in cm, we estimate 
a uniform heating rate of $9.2 \times 10^{-2}$ ergs cm$^{-3}$ s$^{-1}$.

This value is just an initial estimate of the steady uniform heating rate based 
on assumptions, such as constant loop pressure and a simplified radiative loss 
function.  
Using this estimate, we solve the one-dimensional hydrodynamic equations of 
continuity, momentum, and energy with a steady-state solver 
\citep{schrijver2005}.  The radiative loss function we use to solve the 
equations was described by \cite{brooks2006}.  The density and temperature along 
the loop for this heating rate is shown in the bottom two panels in Figure 
\ref{fig:example} as solid lines . 

To calculate the expected intensities in the \ion{Fe}{12} lines from this 
solution, we integrate the emission measure times the emissivity functions 
($\epsilon_\lambda$) of \ion{Fe}{12} 186.880 and 195.119\,\AA, i.e.,
\begin{equation}
I_\lambda = \frac{A}{4\pi}\int_{0}^{s_{1}} \epsilon_\lambda (T,n) n^2 ds.
\label{eqn:intensity}
\end{equation}
In the above equation, $A$ is the area of a pixel (in this case, 1\arcsec) 
and the upper limit of the integration, $s_1$, is the largest distance along the 
loop that still projects into the original footpoint pixel.  In the example, the 
initial guess of the heating rate of $9.2 \times 10^{-2}$ ergs cm$^{-3}$ 
s$^{-1}$ produced a \ion{Fe}{12} 186.880\,\AA\ intensity of 4910 ergs cm$^{-2}$ 
s$^{-1}$ sr$^{-1}$ and a \ion{Fe}{12} 195.119 intensity of 6950 ergs cm$^{-2}$ 
s$^{-1}$ sr$^{-1}$.  

The intensity found in Equation \ref{eqn:intensity} would be the intensity if 
the plasma filled the area of the pixel.  To determine the actual percentage of 
the pixel that contains plasma, or the filling factor, we take the ratio of the 
observed to modeled \ion{Fe}{12} 195.119\,\AA\ intensities.  For this 
simulation, the filling factor is 0.178.  Then we multiply the simulated 
\ion{Fe}{12} 186.880\,\AA\ intensity by this filling factor.  The \ion{Fe}{12} 
186\,\AA\ intensity becomes 873 ergs cm$^{-2}$ s$^{-1}$ sr$^{-1}$ which is 
slightly larger than the observed 819.5 ergs cm$^{-2}$ s$^{-1}$ sr$^{-1}$.

Because the simulated intensity in \ion{Fe}{12} 186.880\,\AA\ is too large, we 
reduce the heating rate by a factor of 2; if it had been too small, we would 
have increased the heating rate by a factor of 2. Using the new heating rate, we 
solve again the one-dimensional hydrodynamic equations. The temperature and 
density solution for a heating rate of $4.6 \times 10^{-2}$ ergs cm$^{-3}$ 
s$^{-1}$ is shown in the lower panels of Figure~\ref{fig:example} with dashed lines.  
This solution produced a \ion{Fe}{12} 186.880\,\AA\ intensity of 755 ergs 
cm$^{-2}$ s$^{-1}$ sr$^{-1}$.   

We now have two different heating rates and two different \ion{Fe}{12} 
186.880\,\AA\ intensities that bracket the desired \ion{Fe}{12} 186.880\,\AA\ 
intensity.  We interpolate between them and predict a heating rate of  $6.8 
\times 10^{-2}$ ergs cm$^{-3}$ s$^{-1}$ will achieve the desired \ion{Fe}{12} 
186.880\,\AA\ intensity.  The temperature and density solution for this heating 
rate is shown in Figure~\ref{fig:example} with dash-dot lines.  The \ion{Fe}{12} 186.880 and 
195.119\,\AA\ intensities determined from Equation~\ref{eqn:intensity} were 1791 
and 2702 ergs cm$^{-2}$ s$^{-1}$ sr$^{-1}$, respectively.  Using the 
\ion{Fe}{12} 195.119\,\AA\ ratio, we determine a filling factor of 0.457.  The 
simulated \ion{Fe}{12} 186.880\,\AA\ intensity is then 819 ergs cm$^{-2}$ 
s$^{-1}$ sr$^{-1}$ which matched the observed intensity.

We complete this process for all field lines in our study.  For 10\% of the 
field lines, the necessary filling factor was $>$ 1; we eliminated these field 
lines from the study.  After the best solution was found for all the field 
lines, a DEM of was generated from the portion of the density and temperature 
that projected into the region of interest (the white rectangle shown in 
Figure~\ref{fig:multipanel}.)  Figure \ref{fig:modeldem} shows the comparison of 
the DEM calculated from the data (black) and the DEM predicted by the steady 
heating model (green).  The model DEM well approximates the observed DEM around its peak, 
but the model DEM does not well describe the observations at lower or higher temperatures.  
We discuss this discrepancy in the next section.

Through this process, we solved for the density and temperature along the loop, 
as well as the required filling factor to bring the \ion{Fe}{12} intensities 
into agreement with observations.  Figure \ref{fig:fftl} shows the resulting 
relationship between the filling factor and temperature (left panel) and loop 
length and temperature (right panel).


\section{Discussion}

In this paper, we have combined \textit{Hinode} EIS and XRT observations to 
calculate a differential emission measure in an active region core over the 
neutral line.  We also measured the densities at the footpoints of the core 
loops using the EIS \ion{Fe}{12} line ratio. Using potential field 
extrapolations of the region we approximate the lengths and geometries of the 
core loops.  Using the density sensitive EIS lines and the loop lengths and 
geometries, we construct a simple steady heating model.  The heating is 
uniformly deposited along the loop and the cross section of the loop is assumed 
to expand as magnetic field decreases.  Using a code that solves the 
one-dimensional hydrodynamic equations for steady heating, we adjust the magnitude 
of the heating rate 
iteratively until the simulated intensities in the \ion{Fe}{12} 186.880 and 
195.119 spectral lines match the observed intensities to within 1\%.  We then construct a 
DEM from the solutions and compare with the observed DEM.

In the temperature range 6.3 $<$ Log T $<$ 6.7, the model DEM is in general 
agreement with the observations.  The magnitude of the model DEM is 
approximately the same as the magnitude of the observed DEM.  The median 
temperature of the model DEM (Log $T = 6.5$ MK) is the same as the temperature 
of the peak of the observed DEM.  This good agreement was arrived at using 
relatively few assumptions: potential field geometry, CHIANTI atomic physics, 
uniform, steady heating, and loop area inversely proportional to the magnetic 
field.  These assumptions were motivated by successful parameter space searches 
in previous studies.   

Unlike previous studies, however, no additional ``fudge factors'' were used, 
such as a filling factor, to force agreement between the observations and the 
model.   Because this study relies on density sensitive line intensities, the 
areal filling factor is determined through the model calculation.  After making 
these four key assumptions, there are no additional assumptions that could be 
made that could, for instance, artificially raise or lower the temperatures in 
the loops.  In fact, the maximum and minimum temperature of model DEM are 
defined by the maximum and minimum density and loop length shown in 
Figures~\ref{fig:density} and \ref{fig:magfield}.  
We determined the maximum electron density in the 
moss was  $\sim 5 \times 10^{10}$ cm$^{-3}$ and the maximum loop length 
measured from potential field extrapolations was $\sim 100$\,Mm.  The moss 
densities were measured with density sensitive Fe~{\sc xii} lines formed at 
$1.5$\,MK. For the resulting pressure ($\sim 20$\,dyn cm$^2$) and half-length, 
we estimate the maximum apex temperature of loops for high-frequency heating is 
6.5\,MK from the RTV scaling law which is in agreement with the maximum 
temperature of the model DEM.    Observations of significant emission
at higher temperatures would have indicated a disagreement with the steady heating model.

\begin{figure*}[t!]
\centerline{
\resizebox{18cm}{!}{\includegraphics{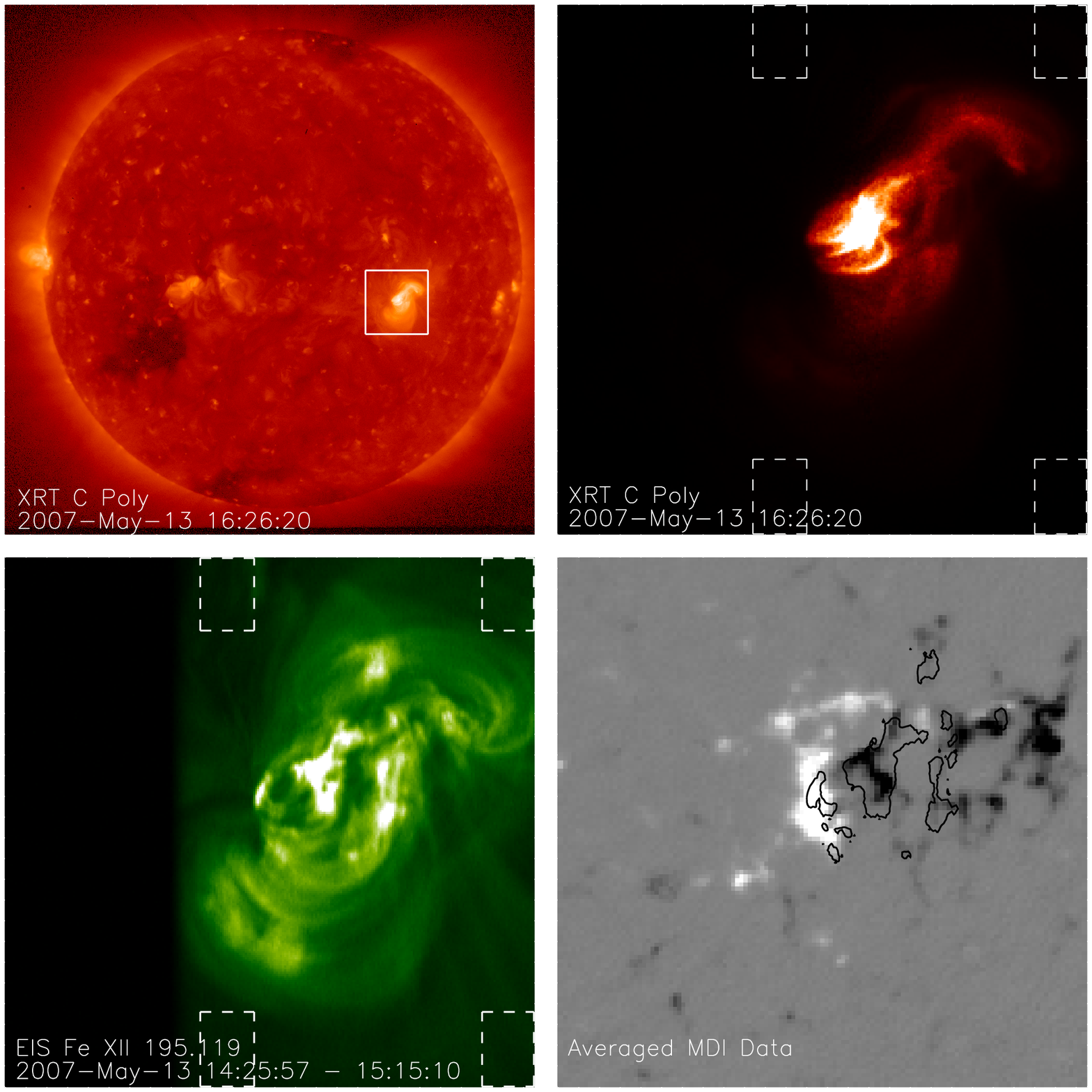}}}
\caption{({\it Upper Left Panel:}) Full disk image of the Sun in XRT C\_poly 
filter 
scaled logarithmically.  The box shows the EIS field of view.  ({\it Upper Right 
Panel:}) A C\_poly image (scaled linearly) in the EIS field of view.  ({\it 
Lower Left 
Panel:}) The EIS \ion{Fe}{12} 195.119\,\AA\ raster of the active region.  The 
squares 
made with dashed lines show the areas used for background subtraction. ({\it 
Lower 
Right Panel:}) The time average photospheric magnetic field measurements from 
MDI in the same field of view as the EIS data.  The contours are from the 
\ion{Fe}{12} 
186.880\,\AA\ raster.
\label{fig:fulldisk}}
\end{figure*}

\begin{figure*}[t!]
\centerline{
\resizebox{18cm}{!}{\includegraphics{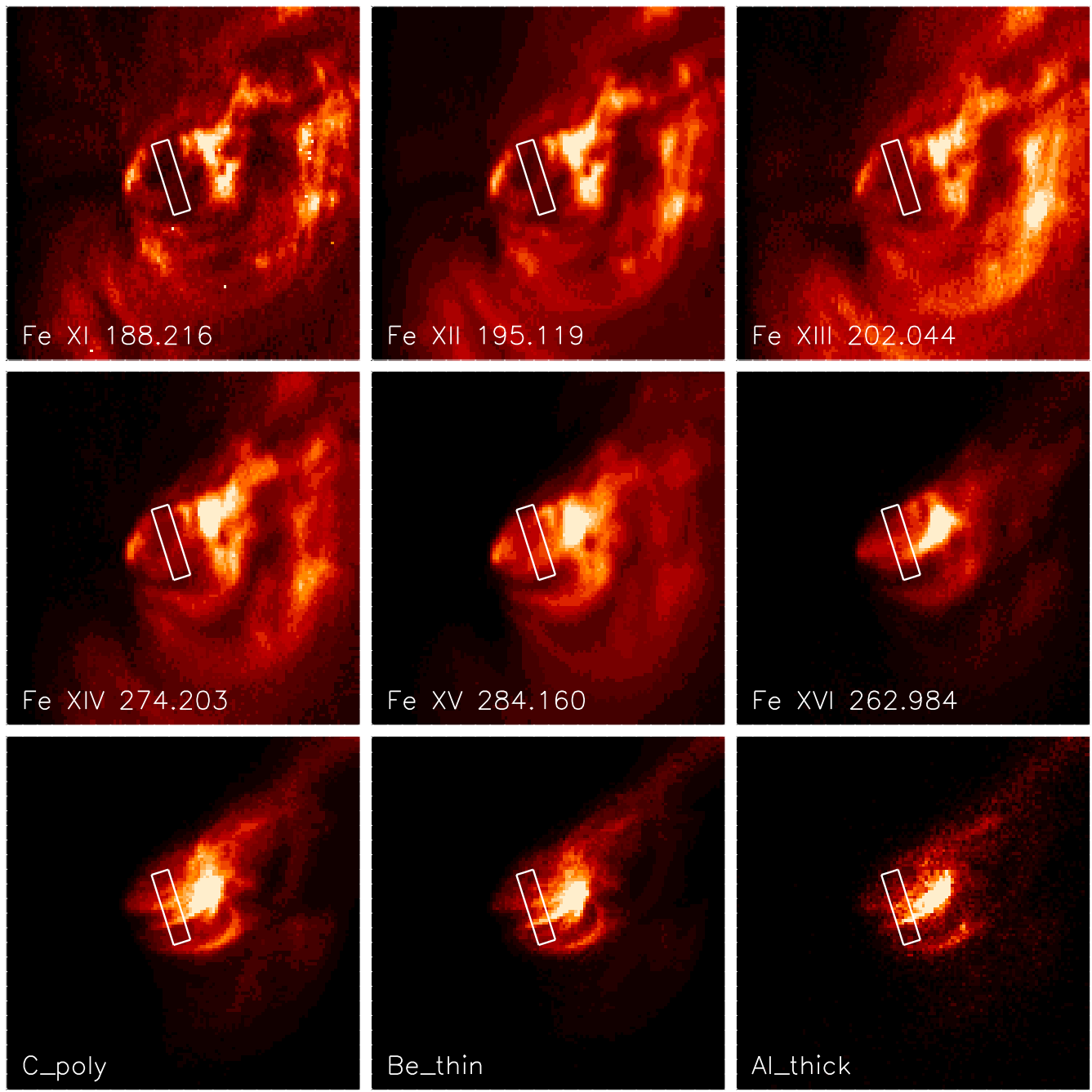}}}
\caption{Several EIS rasters and XRT images from the data set.  The solid lines 
indicate the region over which the intensity was averaged.  
\label{fig:multipanel}}
\end{figure*}

\begin{figure*}[t!]
\centerline{
\resizebox{18cm}{!}{\rotatebox{90}{\includegraphics{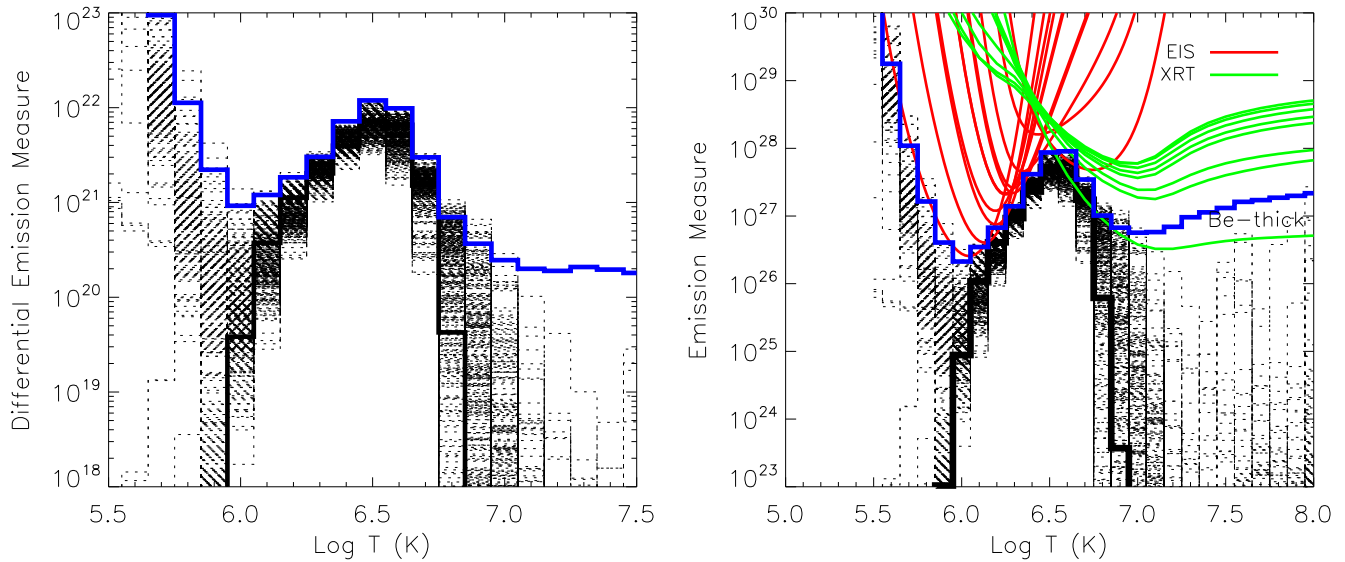}}}}
\caption{({\it Left Panel:}) The thick solid lines show the best DEM for the 
input XRT 
and EIS intensities and the dotted lines surrounding the solid line gives an 
estimate of the error in the DEM. The blue line represents the maximum emission 
that can be added to a single temperature bin without changing the modeled 
intensity by more than the observed errors.  ({\it Right Panel:}) The integral 
form of 
the DEM and corresponding EM loci curves.  The EIS EM loci curves are shown in 
pink and the XRT EM loci curves are shown in green.
\label{fig:dem}}
\end{figure*}

\begin{figure*}[t!]
\centerline{
\resizebox{18cm}{!}{\includegraphics{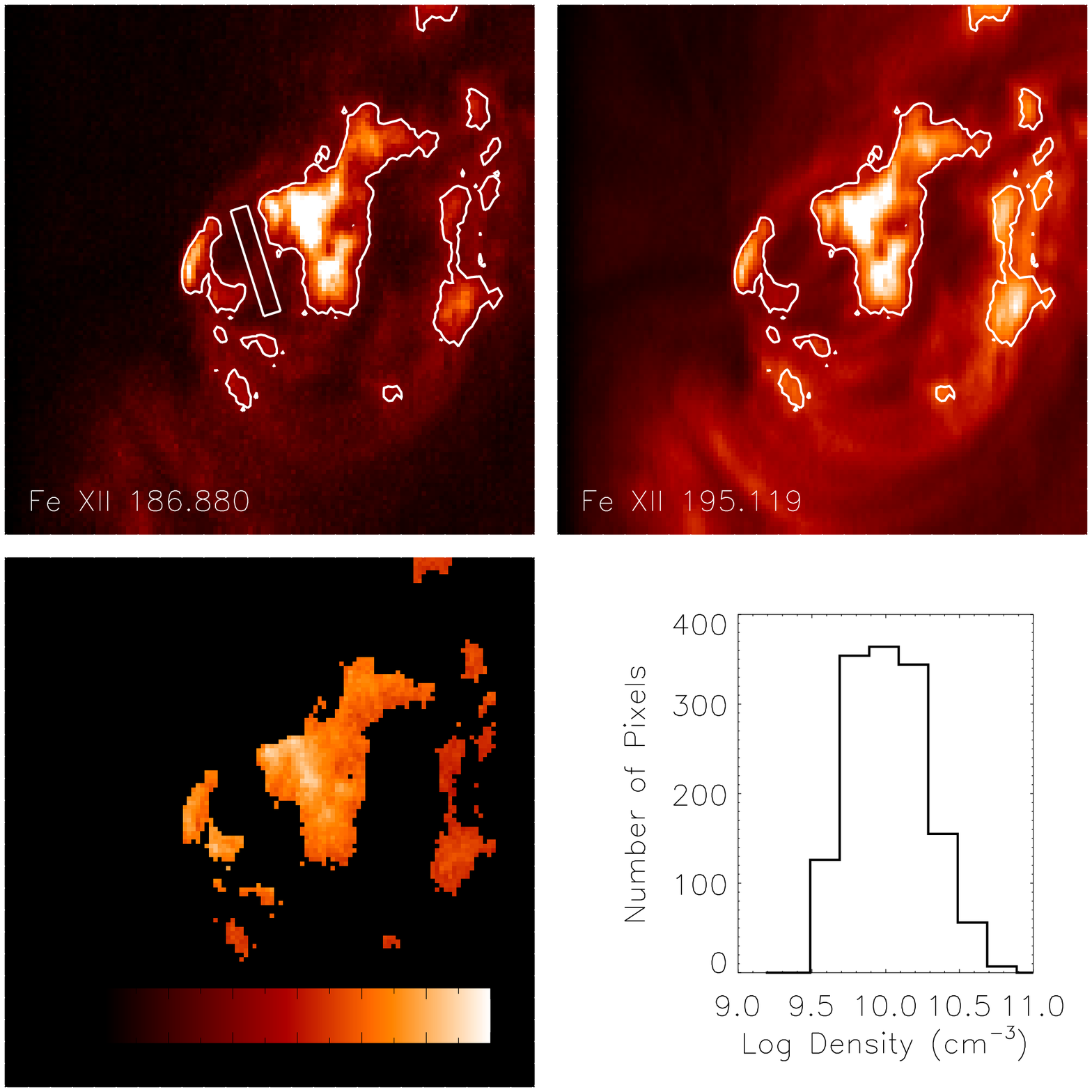}}}
\caption{({\it Upper Panels:}) EIS \ion{Fe}{12} 186.880 and 195.119\,\AA\ 
intensities.  The 
regions we define as moss are shown with a contour.   ({\it Lower Left Panel:}) 
The 
density calculated from the line intensity ratio.  ({\it Lower Right Panel:}) A 
histogram of the densities.  
\label{fig:density}}
\end{figure*}

\begin{figure*}[t!]
\centerline{
\resizebox{18cm}{!}{\includegraphics{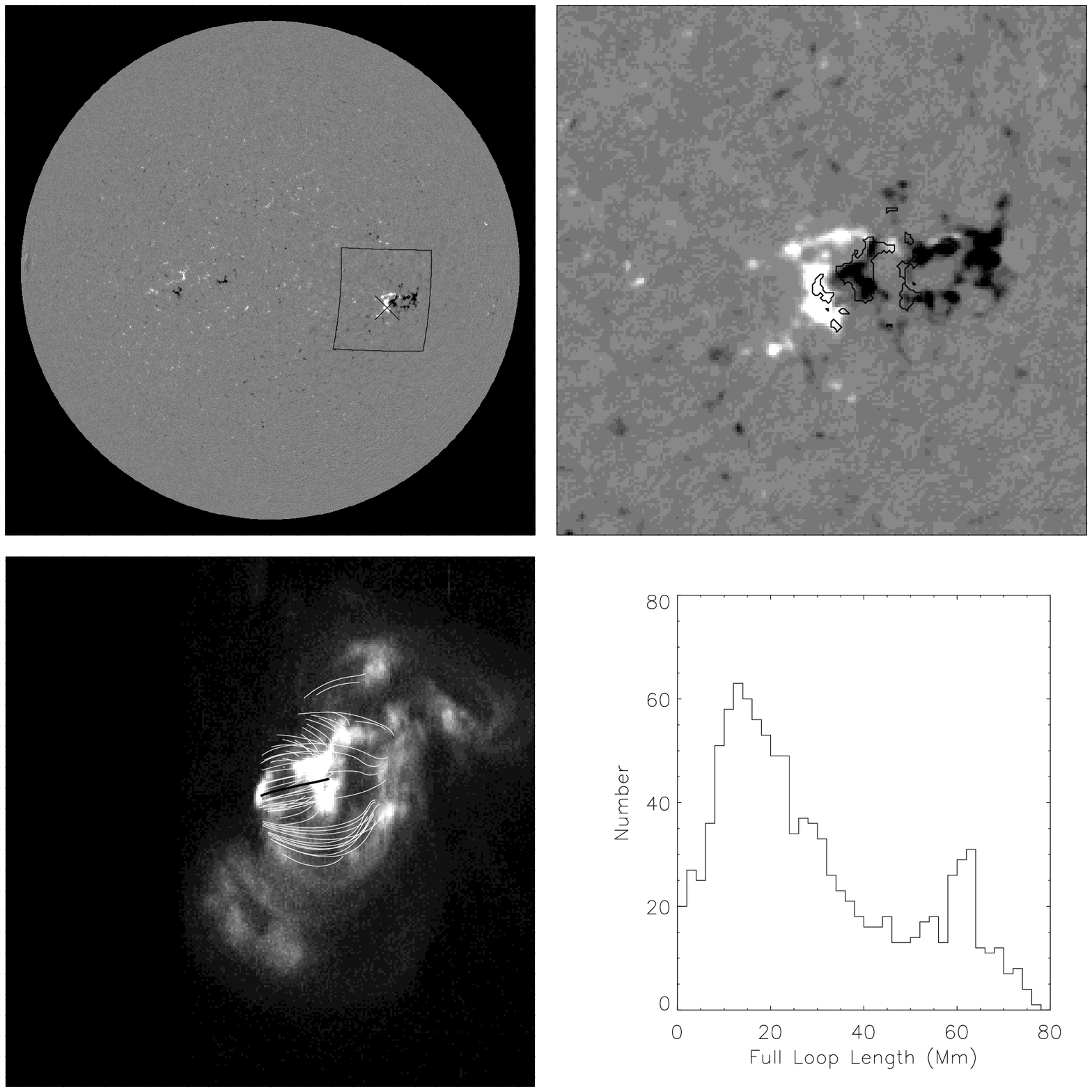}}}
\caption{({\it Upper Left Panel}) Full disk magnetic field observations.  ({\it 
Upper
Right Panel}) Magnetic field around the active region.  The contours are the 
\ion{Fe}{12} 
186.880 moss contours.  ({\it Lower Left Panel}) Field lines from potential 
field 
extrapolation.  ({\it Lower Right Panel}) Histogram of loop lengths of the moss 
loops.
\label{fig:magfield}}
\end{figure*}

\begin{figure*}[t!]
\centerline{
\resizebox{18cm}{!}{\includegraphics{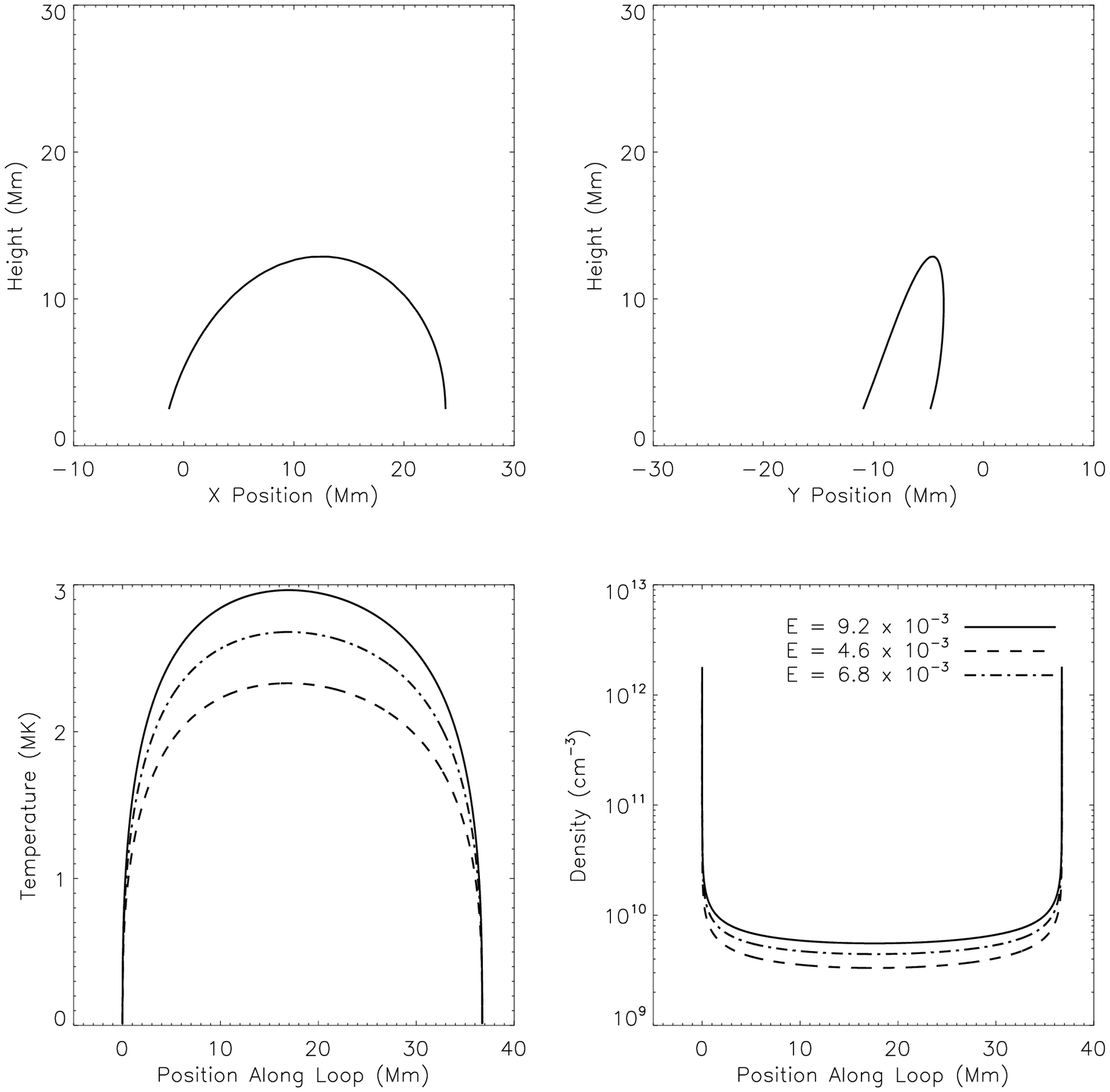}}}
\caption{({\it Upper Panels:}) The field line projected in (x,z) and (y,z).
({\it Lower Left Panel:}) The temperature as a function of position along the 
loop for three different heating rates. ({\it Lower Right Panel:}) The density 
as a function of position along the loop for three different heating rates.
\label{fig:example}}
\end{figure*}

\begin{figure*}[t!]
\centerline{
\resizebox{18cm}{!}{\includegraphics{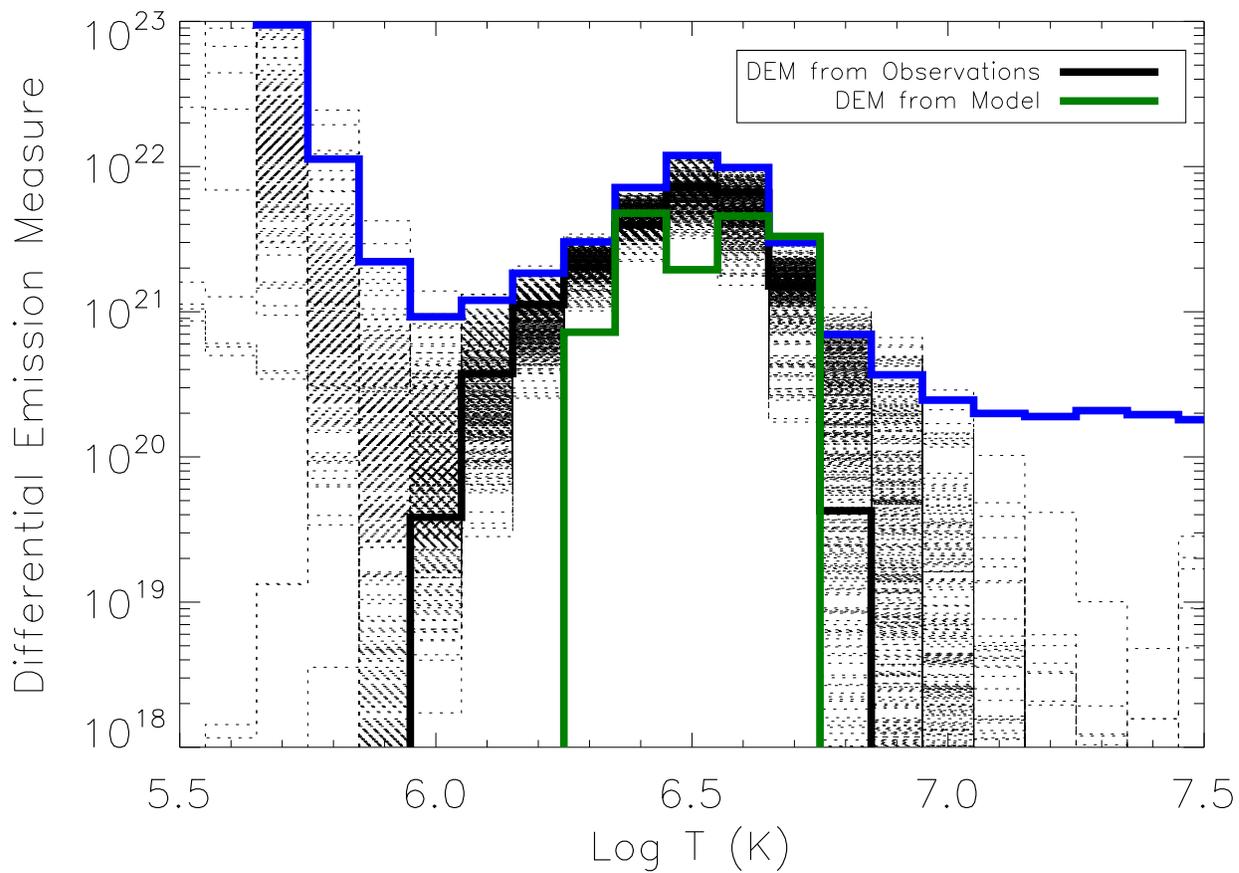}}}
\caption{DEM calculated from XRT and EIS data (black). DEM calculated from a 
steady heating model (green).
\label{fig:modeldem}}
\end{figure*}

\begin{figure*}[t!]
\centerline{
\resizebox{18cm}{!}{\includegraphics{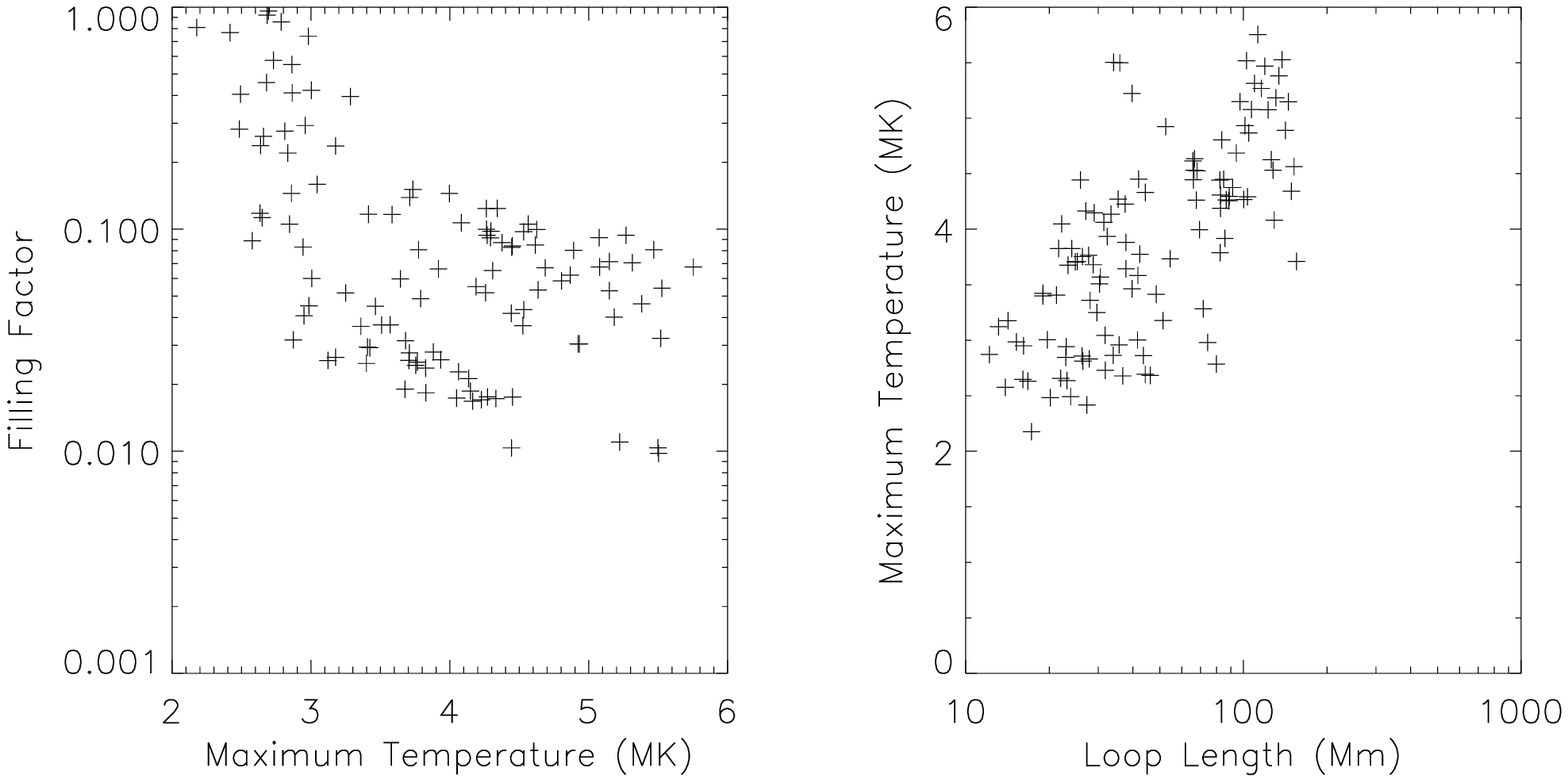}}}
\caption{({\it Left Panel:})  Filling factor as a function of maximum 
temperature in the loop.  ({\it Right Panel:}) Maximum temperature in a loop 
as a function of the full loop length.
\label{fig:fftl}}
\end{figure*}

At temperatures much lower or higher than the peak temperature
(Log T $<$ 6.1 or Log T $>$ 6.7), the observed DEM is not well constrained and 
comparisons with the model are not meaningful.  At mid-range, ``warm'' 
temperatures (6.1 $<$ Log T $<$ 6.3), however, the observed DEM is well 
constrained and significantly larger than the DEM from the model.  There are (at 
least) two possible explanations for this discrepancy. 1) The warm emission is 
from the overlying arcade of warm loops that was not removed in the original 
background subtraction. Figure \ref{fig:fulldisk} supports this view. The 
\ion{Fe}{12} raster shows the presence of many overlying loops that are not 
confined to the core loops bright in the XRT C\_poly image. 2) The warm emission 
is truly from the core loops, casting serious doubt on the steady heating model. 
Another heating model, such as the infrequent heating or ``nanoflare'
model \citep{cargill1997} must be operating. 

The series of TRACE 171\,\AA\ images taken throughout the time period under 
investigation seem to support option (1). The 171\,\AA\ filter has a peak 
response of Log T $\approx$ 6.0, and is extremely sensitive to these warm loops. 
The TRACE images show a series of these loops, with a detailed geometry that 
would remain even after our careful attempts at background subtraction. Based on 
these comparisons, we conclude that the steady heating model agrees with the 
observed properties in the active region core, including the density sensitive 
line ratios in the footpoints of core loops, the distribution of high 
temperature emission at the apex of the core loops, and the appearance of warm 
loops in the series of TRACE images.  Regardless, additional modeling efforts 
are underway to test whether the low-frequency nanoflare model (as described by 
\citealt{cargill1997, cargill2004, klimchuk2008, klimchuk2009}) could be operating 
in active region cores.  Initial studies characterizing the
DEM predicted by long nanoflare storms are underway 
\citep{klimchuk2008,susino2010,tripathi2011,warren2011,mulu2011b}.

In this paper, we chose to use the cooler \ion{Fe}{12} 186.880/195.119\,\AA\ 
line ratio in the moss region and the hotter \ion{Fe}{13} 202.044/203.826\,\AA\ 
line ratio to determine the density in the core region.  It has been well 
documented, however, that densities calculated from the \ion{Fe}{12} line ratio 
could be as much as a factor of 2 more than the densities calculated from other 
EIS line ratios \citep{warren2010a}.  For steady heating, the apex temperature 
of a loop is proportional to the cube root of the base pressure, i.e., $T_{\rm 
apex} \sim (p_0)^{1/3}$, so if the density were a factor of 2 lower, the 
temperatures in the loops would be a factor of 1.25 lower which would shift the 
model DEM to lower temperatures.  Currently, the average median temperature of 
the simulated loops is 3.1 MK. If the true densities were a factor of 2 lower, 
it would shift this average temperature to 2.5 MK.

The densities measured in this analysis are in good agreement with other measurements
of moss densities \citep{tripathi2008,tripathi2010}.  The emission measure curve
found in this analysis, however, is significantly steeper in the temperature
range of 6.0 $<$ Log T $<$ 6.5 than have been found in previous emission measure
calculations.  We find the emission measure curve in this temperature range 
can be approximated as a power-law, i.e., $EM \sim T^b$, 
with an index, $b$, of 3.2, while previous results have found $1 < b < 3$ 
\citep[e.g.,][]{dere1993,brosius1996}.  
These previous results averaged the intensities over large fields of view which included 
both high temperature loops, moss, and extended EUV loops.  Two recent analyses of the
emission measure distribution of inter-moss regions determined a range of indices 
$2.1 < b < 3.4$ \citep{warren2011,tripathi2011}.
It is clear a systematic study of the emission measure
distributions of inter-moss regions is required to fully characterize the temperature
structure of the active region core.

\begin{figure*}[t]
\centerline{
\resizebox{18cm}{!}{\includegraphics{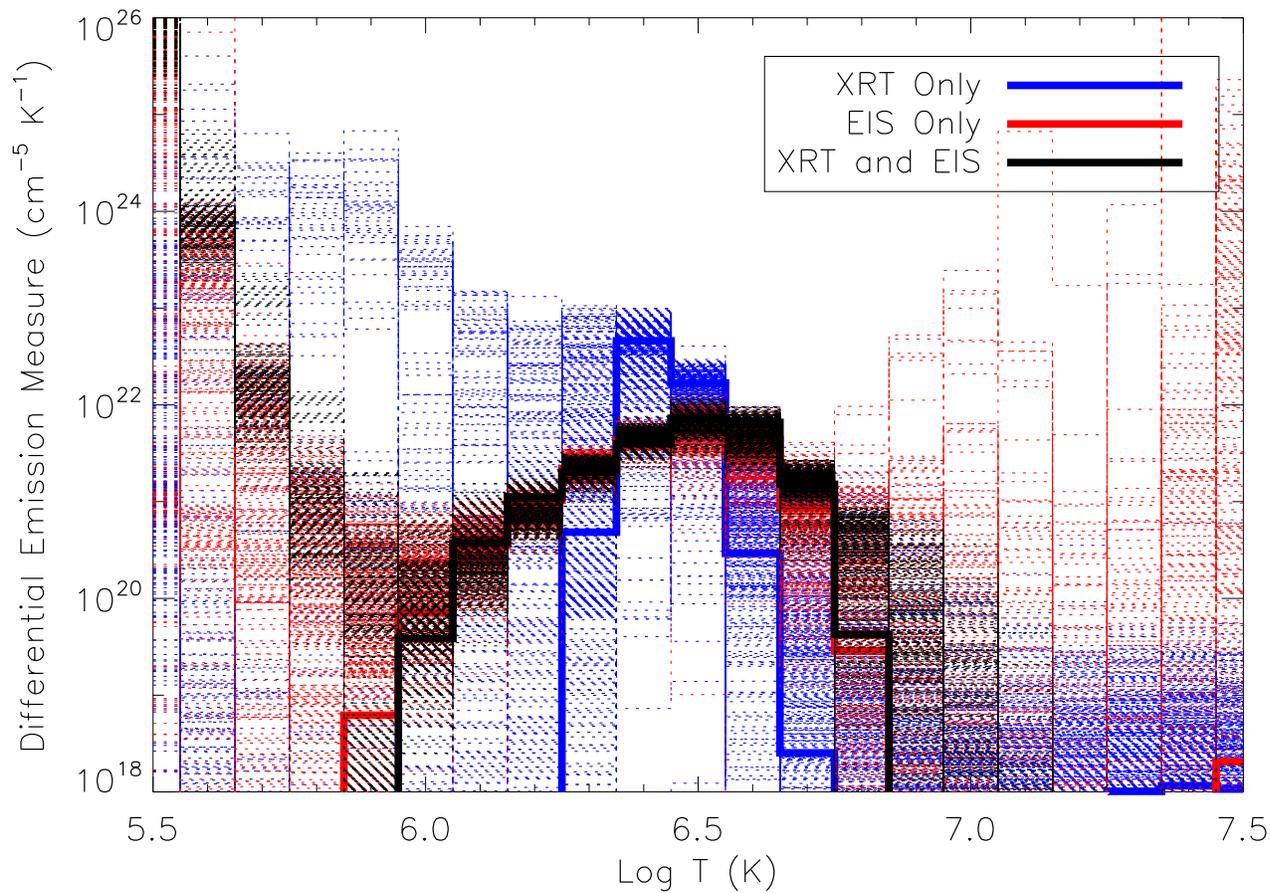}}}
\caption{DEM calculated from XRT intensities only (blue), from EIS intensities 
only (red) and from the combined XRT and EIS data set (black).
\label{fig:dem_all}}
\end{figure*}

In this paper, we combine XRT and EIS data sets to calculate a differential 
emission measure, similar to \cite{schmelz2010}.  Figure \ref{fig:dem_all} 
demonstrates the power of this combined data set compared to using the data sets 
individually.  The DEM shown in blue is calculated from the XRT intensities 
alone, the DEM shown in red is calculated from the EIS line intensities alone, 
and the DEM shown in black is the combined data set.  Comparing the combined DEM 
with the individual instrument DEMs, we see the combined DEM agrees well with 
the EIS DEM in the range 6.0 $<$ Log T $<$ 6.5, however the EIS DEM greatly 
overestimates the true DEM at high temperatures.   The DEM calculated from the 
XRT data is significantly different from the combined DEM at all temperatures.  
The peak of the XRT only DEM is larger and at a lower temperature than the peak 
of the combined DEM.  Because the EIS data highly constrains the combined DEM at 
Log T = 6.5, it forces the emission in the X-ray filters to high temperatures, 
causing the combined DEM to be broader than the XRT only DEM.

We have not considered any cross-calibration factor for these two instruments
in this analysis.  \cite{testa2011} completed an extensive cross-calibration
effort and determined a cross-calibration factor of $\sim 2$ may be necessary.  To
determine the calibration factor, they used EIS lines with significant temperature overlap
of the XRT filters.  The data set presented in this analysis, however,
contained only the \ion{Fe}{17} 254.870\,\AA\ line that overlaps the XRT temperature
range.
Furthermore, additional studies of combined data sets that did have significant 
temperature overlap \citep{warren2011,odwyer2011} did not find the need to include
a cross-calibration factor.  Because the XRT contamination varies
in time, it is difficult to assess the relative calibration of the two instruments.

Finally, our DEM curve for AR 10955, which used XRT data from 16:00 UT (see 
Table 1), shows no significant hot plasma with Log T $\gtrsim$ 7.0; the blue 
line in Figure 3 shows our limited sensitivity to such plasma, given the 
observed count rates and errors in the count rates. This active region, however, 
has been studied previously. \cite{schmelz2009a} used a similar XRT data set 
from 18:00 UT (approximately two hours after the XRT data considered in this 
paper) to determine the DEM for a portion of the active region to the north-west 
of the core. \cite{schmelz2009b} used the 18:00 UT XRT data as well as RHESSI 
upper limits to modify and further constrain the DEM. Both papers found that 
considerable emission in the high temperature range was required to account for 
the small but significant signal detected in the XRT Be\_thick filter.

Although both the 16:18 UT and 18:00 UT Be\_thick images had the same 
resolution, field-of-view, and exposure time, the earlier image analyzed in this 
paper does not show a significant signal, where the later one analyzed by 
\cite{schmelz2009a,schmelz2009b} does. The GOES signal for 2007 May 13 was low 
all day, rarely getting above level A0. There were two small A4 flares from AR 
10955 at 11:20 and 11:35 UT, before both sets of XRT observations, but then the 
region settled down to its sub-A0 level again. It maintained the quiescent stage 
through our 16:00 UT observations, only to rise again to A1-A2 from 17:00 UT to 
just past 18:00 UT when there was another small A7 flare. One possible 
explanation for the different DEM results is that AR 10955 appears to have been 
in a more quiescent state during our 16:00 UT observations, and in a somewhat 
heightened state of activity (right before the small A7 flare) during their 
18:00 UT observations. It would be interesting to repeat the XRT-RHESSI DEM 
analysis for a stronger, more powerful active region with a significant signal 
in the XRT Be\_thick filter to see if the high-temperature plasma is detectable.


\acknowledgments
\textit{Hinode} is a Japanese mission developed and launched by ISAS/JAXA, 
with NAOJ as domestic partner and NASA and STFC (UK) as international 
partners. It is operated by these agencies in co-operation with ESA 
and the NSC (Norway).

ARW was supported by an NSF Career grant.
Solar physics research at the University of Memphis is supported by a 
\textit{Hinode} subcontract from NASA/SAO.  ARW thanks Mark Weber for many 
enlightening conversations on Differential Emission Measures.



\begin{deluxetable}{lllll}
\tabletypesize{\scriptsize}
\tablewidth{0pt}
\tablecaption{ XRT Data from 2007 May 13}
\tablehead{
\colhead{Filter } & \colhead{Time (UT) } & \colhead{Exposure (long) } & 
\colhead{Time (UT) } & \colhead{Exposure (short) }
}
\startdata
Al\_mesh           &16:21:05              &4.10 sec                &16:20:11              
&0.18 sec              \\
C\_poly            &16:26:20              &8.20 sec                &16:25:22              
&0.51 sec              \\
Ti\_poly           &16:19:52              &8.20 sec                &16:19:42              
&0.51 sec              \\
Al\_poly-Ti\_poly  &16:22:55              &16.4 sec                &16:21:50              
&1.45 sec              \\
C\_poly-Ti\_poly   &16:24:38              &16.4 sec                &16:23:37              
&1.03 sec              \\
Be\_thin           &16:28:49              &23.1 sec                &16:28:13              
&1.03 sec              \\
Be\_med            &16:30:23              &46.3 sec                &16:29:32              
&2.05 sec              \\
Al\_thick          &16:18:44              &46.3 sec                &16:18:18              
&16.4 sec              \\
Be\_thick          &16:17:02              &65.5 sec                & ------              
&   -----
\enddata

\label{tab:xrtdata}
\end{deluxetable}

\begin{deluxetable}{|l|ll|ll|}
\tabletypesize{\scriptsize}
\tablewidth{0pt}
\tablecaption{EIS and XRT Intensities}
\tablehead{
\colhead{Line/Filter } & \colhead{I} & \colhead{$\sigma$}  & \colhead{I$_{\rm 
DEM}$} & \colhead{I/I$_{\rm DEM}$}
}
\startdata
\ion{Fe}{10} 177.239  &  254   &  194    &      239  &  1.1   \\
\ion{Fe}{11} 180.401  &  482   &  218    &      637  &  0.76  \\
\ion{Fe}{12} 186.880  &  587   &  149    &      434  &  1.4   \\
\ion{Fe}{12} 195.119  &  768   &  259    &      941  &  0.81  \\
\ion{Fe}{13} 203.826  & 1725   &  414    &     1153  &  1.5   \\
\ion{Fe}{13} 202.044  &  601   &  213    &      421  &  1.4   \\
\ion{Fe}{14} 264.787  & 1016   &  241    &      1189 &  0.85  \\
\ion{Fe}{14} 274.203  &  870   &  215    &       864 &  1.0   \\
\ion{Fe}{15} 284.160  & 9450   & 2150    &     11570 &  0.82  \\
\ion{Fe}{16} 262.984  & 1140   &  257    &       852 &  1.3   \\
\ion{Fe}{17} 254.870  & 33     &  16     &       58 &  0.57   \\
Al\_mesh\tablenotemark{a} & 1940 & 390   &       1406&  1.4   \\
C\_poly      &   1226      &     247     &       1075 & 1.1   \\ 
Ti\_poly     &    876      &     177     &        690 & 1.3   \\
Al\_poly/Ti\_poly &  601   &     121     &        475 & 1.3   \\
C\_poly/Ti\_poly  &  435   &      88     &        385 & 1.1   \\
Be\_thin     &        351  &      71     &        326 & 1.1   \\
Be\_med      &       67.3  &     14.0    &         71 & 0.95  \\
Al\_thick    &       2.43  &     0.82    &       3.23 & 0.75  \\
Be\_thick    &      0.041  &     0.058   &       0.12 & 0.32  \\
\enddata
\label{tab:intensity}
\tablenotetext{a}{The Al\_mesh filter was not used in the DEM calculation.}
\end{deluxetable}

\end{document}